\documentclass[prl,reprint,superscriptaddress,preprintnumbers]{revtex4-1}
\usepackage{amsfonts, amssymb, amsmath}
\usepackage{mathrsfs}
\usepackage{hyperref}
\usepackage[T1]{fontenc}
\usepackage[table]{xcolor}
\usepackage{pifont}

\def\im{{\rm i}}

\def\cF{{\cal F}}

\def\cC{{\cal C}}

\def\cC{{\cal C}}
\def\cH{{\cal H}}

\newcommand{\e}{\mathrm{e}}

\newcommand{\be}{\begin{equation}}
\newcommand{\ee}{\end{equation}}
\newcommand{\bea}{\begin{eqnarray}}
\newcommand{\eea}{\end{eqnarray}}

\newcommand{\?}{\;\!}

\newcommand{\Iprod}[2]{\langle {#1}, {#2} \rangle}

\newcommand{\qv}{\mathcal{R}}
\newcommand{\sv}{\mathcal{I}}
\newcommand{\Jpar}{\nu}

\definecolor{cardinal}{rgb}{0.6,0,0}
\definecolor{darkgreen}{rgb}{0,0.4,0}
\definecolor{purple}{rgb}{0.5, 0, 0.5}
\definecolor{golden}{rgb}{0.92, 0.7, 0}
\definecolor{midnight}{rgb}{0, 0, 0.5}
\definecolor{darkblue}{rgb}{0, 0, 0.8}

\begin{document}

\preprint{CPHT-RR 044 07 2019}

\title{Matter-coupled supersymmetric Kerr-Newman-AdS\texorpdfstring{$_4$}{4} black holes}

\date{\today}

\author{Kiril Hristov}

\affiliation{INRNE, Bulgarian Academy of Sciences, Tsarigradsko Chaussee 72, 1784 Sofia, Bulgaria}

\author{Stefanos Katmadas}

\affiliation{Instituut voor Theoretische Fysica, KU Leuven, Celestijnenlaan 200D, B-3001 Leuven, Belgium}

\author{Chiara Toldo}

\affiliation{CPHT, CNRS, Ecole polytechnique, IP Paris, F-91128 Palaiseau, France}

\begin{abstract}
We present new analytic rotating AdS$_4$ black holes, found as solutions of 4d gauged $\mathcal{N}=2$ supergravity coupled to abelian vector multiplets with a symmetric scalar manifold. These configurations have dyonic charges, preserve two real supercharges and have a smooth limit to the BPS Kerr-Newman-AdS$_4$ black hole. We spell out the solution of the $STU$ model admitting an uplift to M-theory on S$^7$. We identify an entropy function, which upon extremization gives the black hole entropy, to be holographically reproduced by the leading $N$ contribution of the generalized superconformal index of the dual theory.

\end{abstract}

\pacs{}
\keywords{}

\maketitle

\section{Introduction}\label{sec:intro}
The AdS/CFT correspondence provides a natural setting for the statistical interpretation of black hole entropy in terms of a microscopic theory.  The derivation of the Bekenstein-Hawking entropy of four-dimensional Anti-de Sitter (AdS) black holes as a leading contribution to the Witten index of the boundary dual field theory \cite{Benini:2015eyy} opened the way to new computations in this regard (see \cite{Zaffaroni:2019dhb} for a review). The microstate description of rotating black holes is of particular importance, since the black holes in our universe are often spinning close to extremality. Supersymmetry in AdS is compatible with the presence of angular momentum, therefore on the gravity side the first step is to find the relevant black hole solutions.
\begin{table}[h]
	\begin{center}
		\setlength{\tabcolsep}{3pt}
		\begin{tabular}{ c || c | c || c | c ||} \cline{2-5} & \multicolumn{2}{|c||}{AdS$_4\ (\Iprod{G}{\Gamma}=0)$} & \multicolumn{2}{|c||}{mAdS$_4\ (\Iprod{G}{\Gamma}=-1) $} \\ \hline
	\multicolumn{1}{|c||}{type}	 & gravity & $+$ matter & gravity & $+$ matter \\ \hline
		\multicolumn{1}{|c||}{$J=0$} & \cite{Romans:1991nq}\textasteriskcentered & \cite{Duff:1999gh}\textasteriskcentered & \cite{Romans:1991nq}\textasteriskcentered & \cite{Cacciatori:2009iz,Katmadas:2014faa,Halmagyi:2014qza} \\ \hline
		\multicolumn{1}{|c||}{$J\neq 0$} & \cite{Kostelecky:1995ei} & \cite{Cvetic:2005zi}$+$\textcolor{blue}{\bf here} & \cite{Caldarelli:1998hg}\textasteriskcentered & \cite{Hristov:2018spe} \\ \hline
		\end{tabular}
	\end{center}
	\caption{Summary of known supersymmetric AdS$_4$ black holes with spherical topology. An asterisk denotes the absence of a regular horizon, i.e. a naked singularity.}
	\label{tab:sol}
\end{table}

In this paper we extend the supersymmetric Kerr-Newman-AdS$_4$ (KN-AdS$_4$) solutions to supergravity models with vector multiplets. This corresponds to the last missing entry in Table \ref{tab:sol}, summarising all known spherical BPS black holes with AdS$_4$ asymptotics. 
The value of the gauged R-symmetry magnetic flux (given by $\Iprod{G}{\Gamma}$ in the symplectically covariant notation we follow) distinguishes between two types of supersymmetry preserving asymptotics \cite{Hristov:2011ye,Hristov:2011qr}. When the flux vanishes we have the asymptotically AdS$_4$ solution, while the case when the flux is fixed to $-1$ is an example of a particular asymptotically locally AdS space that was dubbed "magnetic AdS" in \cite{Hristov:2011ye}. This dichotomy is well-understood on the AdS boundary for three-dimensional supersymmetric theories \cite{Hristov:2013spa}, where either full superconformal symmetry is preserved ($\Iprod{G}{\Gamma}=0$) or there is only a partial supersymmetry via the so-called topological twist ($\Iprod{G}{\Gamma}=-1$). Static spherical BPS black holes in magnetic AdS$_4$ exist only after coupling gravity to additional matter \cite{Cacciatori:2009iz,Katmadas:2014faa,Halmagyi:2014qza} and these admit rotating generalizations found recently in \cite{Hristov:2018spe}. 
Here instead we are after the other branch of rotating solutions, asymptotic to global AdS$_4$. The KN-AdS$_4$ black holes in Einstein-Maxwell theory with cosmological constant (embeddable in minimal gauged ${\cal N}=2$ supergravity) were found in \cite{Carter:1968rr} and their supersymmetric limit was analyzed in \cite{Kostelecky:1995ei,Caldarelli:1998hg}. Considering the $X^0 X^1$ model of one minimally coupled vector multiplet, \cite{Cvetic:2005zi} and \cite{Chow:2013gba} found electric supersymmetric KN solutions upon taking the BPS limit of thermal black hole solutions (hyperbolic horizons in the same model were found in \cite{Klemm:2011xw}). In the present work we focus from the outset on supersymmetric black holes, constructed through the BPS equations of supergravity coupled to vector multiplets. Assuming that the associated scalar manifold is symmetric, we solve the BPS equations for any such model defined by an arbitrary gauging vector.
From the point of view of holography, the main object of interest is the entropy function, which upon extremization with respect to a set of chemical potentials conjugate to the conserved charges gives the entropy of the newly discovered solutions. We find
\begin{align} \label{eq:entropyfn}
\begin{split}
	S = &-2\? \frac{F (X)}{\omega} - F_I (X) P^I + X^I Q_I \\
&   + \frac{\omega}{2}\?\left( {\cal J} + P^I Q_I \right) + \lambda\? (2\? g_I X^I - \omega - 2 \pi i)\ ,
\end{split}
\end{align}
where the $X^I$ are conjugate to the electric charges $Q_I$, $\omega$ is conjugate to ${\cal J}$, $g_I$ are the FI gauging parameters setting the length scale of AdS$_4$, and the prepotential $F(X)$ and its derivatives $F_I(X)$ are model-dependent functions of the $X^I$. $\lambda$ is a Lagrange multiplier imposing a constraint among the chemical potentials such that upon extremization with respect to the independent set of $\omega, X^I$ one recovers the entropy. In the absence of magnetic charges $P^I$, the above entropy function was introduced in \cite{Choi:2018fdc} and further elaborated in \cite{Cassani:2019mms}, based on the previously known example in \cite{Cvetic:2005zi}. We confirm the conjecture of \cite{Choi:2018fdc} for the full $STU$ model with electric charges and its extension in \eqref{eq:entropyfn} for a new BPS solution to the $X^0 X^1$ model including a magnetic charge. The Legendre transform of the entropy function presented above is expected to match the saddle point evaluation of the partition function of the holographically dual theory on (Euclidean) S$^1 \times$S$^2$, which in this case is the generalized superconformal index \cite{Kim:2009wb,Imamura:2011su,Kapustin:2011jm}.

\section{The real formulation of supergravity}\label{sec:formalism}
Our starting point is the action for abelian gauged ${\cal N} = 2$ supergravity with $n_V$ vector multiplets. Our conventions and initial steps coincide with those followed in \cite[Sec.~2]{Hristov:2018spe}. The bosonic fields are the metric $g_{\mu\nu}$, $(n_V+1)$ abelian gauge fields $A^I_\mu (I = 0, .., n_V)$  and $n_V$ complex scalars $z^i (i = 1, ..., n_V)$. The Lagrangian and supersymmetry transformation rules are uniquely specified by a choice of the so-called prepotential $F (X^I)$ and the symplectic vector of Fayet-Iliopoulos (FI) parameters $G = \{g^I, g_I \}$ defining a combination of abelian gauge fields gauging the R-symmetry. 

The BPS equations for solutions with a timelike Killing vector were given in \cite{Cacciatori:2008ek, Meessen:2012sr, Chimento:2015rra}, with a metric 
\begin{equation}\label{eq:metr-bps}
{\rm d} s^2_4 = -\e^{2 U} ({\rm d} t+\omega \? {\rm d}\phi)^2 + \e^{-2 U} {\rm d} s^2_3 \,,
\end{equation}
where ${\rm d} s^2_3$ is the metric of a three-dimensional base space, on which all quantities are defined.

We express the original complex scalars $z^i$ and scale factor $\e^{U}$ in terms of a suitably gauge fixed symplectic section 
\begin{equation}
\e^{-2 U} \qv + \im\? \sv = \{X^I, F_I \} \,,
\end{equation}
where $F_I \equiv \partial F/ \partial X^I$.
We choose to only involve ${\cal I}$ in our explicit ansatz, noting that one can further use 
\begin{equation}\label{eq:r-i}
 \qv = - \frac1{2 I_4(\sv)} I^\prime_4(\sv) = - \frac12 \e^{4 U} I^\prime_4(\sv)\,,
\end{equation}
and then by the special choice of coordinates $z^i = ({\cal R} + i {\cal I})^i / ({\cal R} + i {\cal I})^0$ recover the physical scalars. In writing \eqref{eq:r-i} we already assumed that the special K\"{a}hler manifold parametrized by the scalar fields is a symmetric space, such that we can use the quartic invariant formalism reviewed in \cite[Sec.~2.2]{Hristov:2018spe}. The quartic form $I_4$ is invariant under symplectic transformations, while its derivative $I_4'$ is a symplectic vector and is therefore covariant. One can explicitly evaluate $I_4$ and its derivatives for any given symmetric model. We are especially interested in the so-called magnetic $STU$ model,
\begin{equation}\label{eq:mSTU-def}
F^{mSTU}= 2 i \sqrt{X^0 X^1 X^2 X^3}\,,
\end{equation}
and purely electric gauging $G = \{0, g_I\}$, because the resulting Lagrangian can be embedded in 11d supergravity compactified on S$^7$ \cite{Duff:1999gh,Cvetic:1999xp}. The $I_4$ invariant in this case was explicitly spelled out in \cite[Sec.~2.2]{Hristov:2018spe}.

\section{Base space ansatz}\label{sec:base}
The BPS equations of \cite{Cacciatori:2008ek, Meessen:2012sr, Chimento:2015rra} are conveniently cast using the following metric on the $3d$ base 
\begin{equation}
	{\rm d} s^2_3 = {\rm d} \rho^2 + \e^{2 \varphi} ({\rm d} x^2 + {\rm d} y^2)\ ,
\end{equation}
for a general function $\varphi(\rho, x, y)$. For stationary black hole solutions, one further assumes that $\partial/\partial y$ is also an isometry, leaving us with $\varphi(\rho, x)$. There are two major classes of black hole solutions in the literature, depending on the separability of $\varphi$. The restricted choice 
\be
\e^{2 \varphi}_{CK} = \Phi(x)\ \e^{2 \psi(\rho)}
\ee
leads to Cacciatori-Klemm-type solutions \cite{Cacciatori:2009iz,Katmadas:2014faa,Halmagyi:2014qza} and their rotating generalizations \cite{Hristov:2018spe}. Here instead we focus on another class, leading to Kerr-Newman type solutions \cite{Klemm:2013eca} (and even more generally to the Plebanski-Damianski solution). In this case $\e^{2 \varphi}$ is separable in terms of new coordinates $q$ and $p$, such that
\begin{equation}
	\e^{2 \varphi} = Q(q) P(p)\ , \quad \rho = q\ p\ , \quad x = \alpha(q) + \beta(p)\ ,
\end{equation} 
with arbitrary functions $Q(q), P(p)$, while the functions $\alpha(q)$ and $\beta(p)$ are conventionally chosen as
\begin{equation}
 \alpha^\prime(q) = - \frac{q}{Q(q)}\,, \qquad  \beta^\prime(p) = \frac{p}{P(p)}\,,
\end{equation}
in order to bring the base metric in the diagonal form
\begin{equation}\label{eq:3d-base}
 {\rm d} s^2_3 = \e^{2 \sigma} \left( \frac{{\rm d} p^2}{P(p)}  + \frac{{\rm d} q^2}{Q(q)} \right) + Q(q) P(p) {\rm d} y^2 \,,
\end{equation}
where we defined
\begin{equation}\label{eq:e2sig-def}
 \e^{2 \sigma} \equiv q^2 P(p) + p^2 Q(q) \,.
\end{equation}
The standard form of the base metric for supersymmetric Kerr--Newman is reached upon setting $\{ q, p, y \} \sim \{r, \cos\theta, \phi \}$
where $r$ is a radial coordinate and $\theta$, $\phi$ are coordinates on a sphere.

\section{BPS equations}\label{sec:BPS}
We can massage the set of equations in \cite[Sec.~2.3]{Hristov:2018spe} using the ansatz for the base space discussed above. The symmetries imposed guarantee that the superalgebra of the resulting solution is given by $U(1|1)$, i.e.\ a quarter-BPS configuration. 
We find that the vielbein BPS equation (2.34) in \cite{Hristov:2018spe} for the choice $\hat{G} = G\?{\rm d}(q\?p)$ leads to
\begin{align}\label{eq:struct-base}
 \Iprod{G}{\sv} &= \frac14 \, \e^{-2 \sigma}\frac{ \partial^2 \e^{2 \sigma}}{\partial q \partial p }\,, 
 \\ 
 \Iprod{G}{ {\cal A}} &= \frac12\,\e^{-2 \sigma}(p\, Q(q)\, P^\prime(p) - q\, P(p)\, Q^\prime(q) )\,  {\rm d} y\,, \nonumber
\end{align}
where ${\cal A}$ is the spatial part of the symplectic vector of electric/magnetic vector fields.
The equation for the rotation one-form $\omega$ can be compactly written as
\begin{equation}
  \star {\rm d} \omega =  \Iprod{d \sv}{\sv} + \Iprod{G}{I_4'(\sv)} \?{\rm d}(q\?p) \,.  \label{eq:omega-eqn-gen}
\end{equation}
Finally, we need to solve the BPS equation for the scalars and electromagnetic fields, 
\begin{align}\label{eq:fl-Ortin-s}
\begin{split}
\cF &= q\ p\ G\ {\rm d} \omega -\star\ {\rm d} \sv \\
& - \star {\rm d}(q\?p) \? \left( \Iprod{G}{\sv}\,\sv 
-\frac14  I^\prime_4(\sv , \sv , G) \right)\ .
\end{split}
\end{align}
The equations of motion in this case are implied by the above equations and the requirement that the symplectic vector ${\cal F}$ of field strengths be closed, ${\rm d} {\cal F} = 0$. 
Similar to \cite{Hristov:2018spe}, a rescaling of the symplectic section is convenient for expressing the BPS equations. For the class of solutions based on \eqref{eq:3d-base}-\eqref{eq:e2sig-def} we use
\begin{equation}\label{eq:H-def}
 \cH = \e^{2 \sigma} \?\sv\,, \qquad  I_4(\cH) = \e^{8 \sigma}\e^{- 4U}\,.
\end{equation}
This change of variable in \eqref{eq:struct-base}-\eqref{eq:fl-Ortin-s} brings the BPS equations to a form that can be solved in terms of polynomial ans\"{a}tze for the variables $\e^{2 \sigma}$ and $\cH$, as will be shown explicitly below.

\section{Near-horizon solution}\label{sec:nearhor}
We first want to solve the BPS equations \eqref{eq:struct-base}-\eqref{eq:fl-Ortin-s} near the horizon. We impose an ansatz compatible with the $SU(1,1)$ isometry group of AdS$_2$ such that the superalgebra is further enhanced to $SU(1,1|1)$, a half-BPS configuration. We choose here $q \equiv r$ and the function $Q(r)$ as
\begin{equation}
 Q(r) = R_0^2 r^2 \,,
\end{equation}
where $r$ is a radial coordinate and $R_0$ is a constant, so that $\e^{2 \sigma}$ is also separable
\begin{equation}\label{eq:horsigma}
 \e^{2 \sigma} = r^2 \e^{2 \sigma_0} \,,  \qquad  \e^{2 \sigma_0} =  P(p) + R_0^2 p^2 \,.
\end{equation}
The conical structure of the base space implies the scaling behaviour
\begin{equation}\label{eq:eU-attr}
 \e^{-2  U} = \frac{1}{r^2} \e^{-2 U_0}\,, \quad \omega = \frac{1}{r} \omega_0 \,, \quad  \cH = r\, \cH_0\,,
\end{equation}
for the components of the metric and the scalars, where the functions $U_0, \omega_0$ and $\cH_0$ depend only on $p$. 

With this ansatz, we are left with solving the BPS equations in order to determine the dependence on the coordinate $p$, i.e.\ along the sphere. We  first combine \eqref{eq:omega-eqn-gen} and \eqref{eq:fl-Ortin-s} to solve for the rotation one-form as
\begin{equation}\label{eq:hor-omega}
 \omega_0 = \Jpar\? P(p)\? \e^{-2 \sigma_0} = \Jpar \frac{P(p)}{ P(p) + R_0^2 p^2 }\,,
\end{equation}
where $\Jpar$ is a constant to be fixed in due course. 
The remaining equations can be written in terms of $\e^{2\?\sigma_0}$, the symplectic vector $\cH_0$ and its contractions with the vector of parameters $G$. A polynomial ansatz for $\cH_0$
\begin{equation}\label{eq:H0-ansatz-hor}
 \cH_0 = \cC_3\? p^3 + \cC_2 \? p^2 + \cC_1 \? p + \cC_0 \,,
\end{equation}
allows to integrate \eqref{eq:struct-base} for $\e^{\sigma_0}$ as
\begin{align}\label{eq:solhorsigma}
 \e^{2\?\sigma_0} = &\, \frac{1}{2}\?\Iprod{G}{\cC_3}\? p^4 + \frac{2}{3}\?\Iprod{G}{\cC_2}\? p^3 
   \nonumber\\
   &\, +\Iprod{G}{\cC_1}\? p^2 + 2\?\Iprod{G}{\cC_0}\? p + \Xi^{-1}\,,
\end{align}
with $\Xi$ an integration constant. The BPS equations \eqref{eq:fl-Ortin-s} are then solved order by order in $p$, remaining with a single constant symplectic vector ${\cal C}$,
\begin{align}\label{eq:C-sols-hor}
\cC_0 = &\, \frac{1}{\Xi} \? \cC\ , \qquad  \cC_3 = \frac{1}{2\,\Xi}\? I_4(\cC) \? I_4^\prime(G) \,. \nonumber\\
 \cC_1 = &\, \frac{1}{\Xi}\? \left(\Iprod{G}{\cC}\?\cC + \frac{1}{4}\?I_4^\prime(\cC, \cC, G) \right) \,,\\
 \cC_2 = &\, -\frac{1}{2\,\Xi}\? \left( \Iprod{G}{I_4^\prime(\cC)}\? G - \frac{1}{4}\?I_4^\prime(I_4^\prime(\cC), G, G) \right) \,. \nonumber
\end{align}
The rotation parameter $\Jpar$ is fixed in terms of ${\cal C}$ from
\begin{equation}\label{eq:J-sol}
 \Jpar = -\frac{1}{2\,\Xi}\? \Iprod{G}{I_4^\prime(\cC)} \,.
\end{equation}
Finally, the gauge field strengths are given by
\begin{equation}\label{eq:fieldstr}
	{\cal F} = R_0^2\? {\rm d} \left( \e^{-2\?\sigma_0} p\? \left( \cH_0 - \Jpar\?  p^{2}\? G \right) \? {\rm d} y  \right) \,,
\end{equation}
and can be seen to satisfy automatically the second condition in \eqref{eq:struct-base}. 
 
We have presented a complete supersymmetric solution, which in general may allow for various horizon topologies and features non-vanishing NUT charge. Although such solutions are interesting in their own right, here we focus on compact horizons, requiring that the space spanned by $p$ and $y$ is of spherical topology without any NUT charge. This translates to conditions on the function $P(p)$, as discussed in detail in \cite[Sec.~2.5]{Gnecchi:2013mja}, in particular that $P(p)$ must have two roots and be an even function, which by \eqref{eq:horsigma} and \eqref{eq:solhorsigma} imply
\begin{align}
\label{eq:NUT-hor}
	 \Iprod{G}{\cC} &= 0\ , \qquad \Iprod{I_4^\prime(G)}{I_4^\prime(\cC)} = 0\,,   \\
\label{eq:R0value} 	\Xi\? R_0^2 &= 1+I_4(G)\ I_4(\cC) +\frac{1}{4} I_4 (\cC,\cC,G,G)\ ,
\end{align}
leading to 
\begin{equation}\label{eq:Ptopology}
	 P(p) = \frac{1}{\Xi}\left( 1 - I_4(G)\ I_4(\cC)\ p^2  \right)\ (1-p^2)\,.
\end{equation}
We also need to restrict the range of the coordinate $p$ to only reach until the smaller of the two double roots, and thus arrive at the coordinate redefinition
\begin{equation}
	p = \cos \theta\ , \qquad y = \phi\ ,
\end{equation}
which brings the metric in a more conventional form in terms of the spherical coordinates $\{\theta, \phi \}$, upon the additional requirement that $I_4(G)\ I_4(\cC) < 1$.

The final constraint on the spherical part of the metric comes from the requirement that near the poles $p = \pm 1 (\theta = 0, \pi)$ we recover flat space (i.e.\ no conical singularities), which fixes
\begin{align}
	\Xi = 1-I_4(G)\ I_4 (\cC) \ . 
\end{align}
To make the relation with previous literature more manifest, we can define
\begin{equation}\label{eq:Xi-a}
	\Xi \equiv (1 - \frac{a^2}{l^2})  , \quad a \equiv  \frac{\sqrt{I_4(\cC)}}{l}\ , 
\end{equation}
where $l = (I_4 (G))^{-1/4}$ sets the AdS$_4$ radius, as we show in the next section. The constraint that $\Xi > 0$ also translates in the more familiar $a < l$ and the metric function $P$ becomes
\begin{equation}\label{eq:Poftheta}
	P(\theta) = \frac{\sin^2 \theta}{\Xi}\ \left(1 - \frac{a^2}{l^2} \cos^2 \theta \right)\ .
\end{equation}
The resulting charge vector (with the usual periodicity of $\theta, \phi$) is then computed through \eqref{eq:fieldstr}, as
\begin{equation}\label{eq:attractoreq}
 \Gamma  \equiv \frac{1}{4\?\pi}\?\int \cF  = \frac{1}{\Xi} \left(\cC + \frac{1}{8}\?I_4^\prime(I_4^\prime(\cC), G, G) \right) \,.
\end{equation}
This constitutes the main attractor equation, through which the scalars and metric functions encoded in the vector $\cC$ can be solved for in terms of the charges $\Gamma = \{P^I, Q_I\}$. If we now contract \eqref{eq:attractoreq} with the vector $G$ and use the constraints \eqref{eq:NUT-hor}, we find the anticipated constraint on the magnetic flux of the R-symmetry,
\begin{equation}\label{eq:Rflux}
	\Iprod{G}{\Gamma} = 0\ .
\end{equation}

We can also present the Bekenstein-Hawking entropy in the compact form
\begin{align}\label{eq:Bek-Haw-entropy}
	S =& \frac{\pi}{\Xi}\left(  \Xi\? R_0^2 \? I_4(\cC) - \frac{1}{4} \Iprod{G}{I_4'(\cC)}^2 \right)^{1/2} \ ,
\end{align}
where $R_0$ is given by \eqref{eq:R0value} and we use units where the Newton constant is fixed as $G_N = 1$.

\section{Full flow}\label{sec:fullflow}
We now use $q=r$ as a radial variable that runs between the horizon and the asymptotic AdS$_4$ spacetime. It is natural to extend the near-horizon behaviour of $\cH$ in \eqref{eq:eU-attr} to the more general polynomial ansatz 
\begin{equation}\label{eq:H-flow-ans}
 \cH = r\? \left( \cH_0(p) + \left( \cH_1^{(0)} + \cH_1^{(1)} p \right)\?r + \cH_2^{(1)} p \? r^2 \right)\,,
\end{equation}
where all vectors $\cH_{1}^{(0,1)}$ and $\cH_{2}^{(1)}$ are constant, while $\cH_0$ is automatically identified with the one on the horizon. The highest power of $r$ is dictated by the fact that the $\e^{2\?\sigma}$ arising from \eqref{eq:struct-base} needs to be quartic in $r$ to keep the desired AdS$_4$ asymptotics. With this ansatz the expression for $\e^{2\sigma}$ is
\begin{equation}
 \e^{2\?\sigma} = r^2\?\left( \e^{2\?\sigma_0}  + \frac{2}{3}\?\Iprod{G}{\cH_1^{(1)}\?r + \cH_2^{(1)}\?r^2 }\?p^2  \right)\ ,
\end{equation}
upon disregarding integration constants and imposing $\Iprod{G}{\cH_1^{(0)}}  = 0\,,$ in order to keep the structure assumed in \eqref{eq:e2sig-def}. We also make the following ansatz for the rotation form $\omega$ 
\begin{equation}\label{eq:omega-full}
 \omega = \e^{-2\?\sigma}\left( \mu\?Q(r) + \Jpar\?r\?P(p)  \right) - \mu\,,
\end{equation}
where $\Jpar$ was fixed already on the horizon in \eqref{eq:J-sol}, while $\mu$ is another integration constant.

Plugging in the full flow ansatz for $\cH$ and $\omega$ in the BPS equations \eqref{eq:struct-base}-\eqref{eq:fl-Ortin-s} results in an overconstrained system of equations from the various powers of $p$ and $r$. The solution is eventually fully fixed in terms of the vectors $\cC$ and $G$. We find
\begin{align}
\cH_2^{(1)}= \frac{l^2}{2\, \Xi}\? I^\prime_4(G) \,, \qquad  \cH_1^{(0)} = \frac{l}{\Xi}\? G \,,  \nonumber\\
\cH_1^{(1)} = \frac{l}{4\?\Xi}\  I^\prime_4(\cC, G, G)\,, \qquad \mu = - \frac{l}{\Xi} \,.
\end{align}
The metric function $Q$ therefore becomes
\begin{equation}\label{eq:Qofr}
	Q(r) = \frac{r^2}{\Xi} \left( \Xi\?R_0^2  + l\? \Iprod{\cC}{I_4'(G)}\? r + \frac{r^2}{l^2} \right)\ ,
\end{equation}
where the first term in the bracket also depends explicitly on the vectors $\cC$ and $G$ via \eqref{eq:R0value}. The solutions found here asymptote to AdS$_4$ with boundary metric
\begin{align} \label{eq:bound-metr}
ds^2= \frac{r^2}{\Xi} &\left[- \frac{P(\theta)}{\sin^2 \theta}\?\frac{\Xi}{l^2}\?dt^2 +\frac{\sin^2 \theta}{P(\theta)}\? d \theta^2 \right.
\nonumber \\
& \qquad\qquad \left. + \sin^2 \theta \left( d \phi +  \frac{1}{l} dt\right)^2  \right]\,.
\end{align}
The subleading terms of the metric near this boundary encode the mass $M$ which we computed via the AMD procedure \cite{Ashtekar:1984zz,Ashtekar:1999jx}, and the angular momentum ${\cal J}$, 
 \begin{equation}\label{eq:J-gen}
 {\cal J} = \frac{1}{\Xi}\left( \frac{I_4(\cC)}{\Xi}\?\Iprod{\cC}{I_4'(G)} -  \left( 1 + \tfrac{I_4(\cC)}{l^4} \right) \?\nu \right) \,,
\end{equation}
computed through the Komar integral. Combined with the charges in \eqref{eq:attractoreq}, we find that the following BPS bound is obeyed: 
\begin{equation}\label{eq:bps-stu}
	M = \frac{|{\cal J}|}{l} + \frac{l^3}{2} \, \big|\Iprod{\Gamma}{I_4'(G)}\big|  =  \frac{|{\cal J}|}{l} + \frac{1}{\sqrt{2}} \,  \big| \sum_{I=1}^4 Q_I  \big|\ ,
\end{equation}
where the second term corresponds to the R-symmetry charge $T$ \cite{Hristov:2011ye,Hristov:2011qr} and in the second equality we evaluated explicitly for the $STU$ model below. Finally, the roots of $Q(r)$ in \eqref{eq:Qofr} determine the location of the four horizons. The product of their areas, as expected, depends only on quantized charges \cite{Cvetic:2010mn,Castro:2012av,Toldo:2012ec}:
\begin{equation}
\prod_{\alpha=1}^4 A_{\alpha} = (4 \pi)^4 l_{AdS}^4 \left( I_4(\Gamma) + \mathcal{J}^2 \right)\,.
\end{equation}

\section{Solutions of the STU model}\label{sec:stusol}
We now look for explicit solutions of the $STU$ model. We work in the standard electric gauging frame, with the prepotential \eqref{eq:mSTU-def} and the FI terms given by
\begin{align}\label{eq:STUmodel}
% 	F^{mSTU} = & - 2 i\? \sqrt{X^0 X^1 X^2 X^3}\ , \nonumber\\ 
	G =& \, \{0, 0, 0, 0; g, g, g, g \}\ ,
\end{align}
resulting in AdS$_4$ length scale $l = (\sqrt{2} g)^{-1}$. To write down a solution, we need to find a symplectic vector $\cC$ that satisfies the constraints \eqref{eq:NUT-hor}, leaving us with up to six free parameters: four independent electric and two independent magnetic charges. These constraints are however nonlinear and for the sake of brevity we choose to give a configuration corresponding to four electric and only one independent magnetic charge,
\begin{equation}\label{eq:C-STU}
	\cC^{\scriptscriptstyle STU} = \{-\alpha, \alpha, -\alpha \tfrac{(\beta_0 - \beta_1)}{(\beta_2 - \beta_3)}, \alpha \tfrac{(\beta_0 - \beta_1)}{(\beta_2 - \beta_3)} \,;\,\, \beta_I \}\ ,
\end{equation}
for constant $\alpha$ and $\beta_I$ for $I=1\dots 4$. The vector $\cC$ then determines the conserved charges (through \eqref{eq:attractoreq}) and physical properties of the black holes. The full expressions for the conserved charges are in general rather long, therefore in the following we restrict the parameters $\alpha, \beta_I$ in two different ways that provide a more accessible insight into the properties of the solutions.

\subsection{The \texorpdfstring{$T^3$}{T3} model with electric charges}\label{sec:t3sol}
The $T^3$ truncation is achieved by setting equal the three different vector multiplets. In this case, \eqref{eq:NUT-hor} do not allow for an independent magnetic charge, so we concentrate on the purely electric case, setting $\alpha = 0, \beta_1 = \beta_2 = \beta_3$ in $\cC^{STU}$ to find
\begin{equation}\label{eq:C-T3}
	\cC^{T^3} = \{0,0,0,0; \beta_0, \beta_1, \beta_1, \beta_1 \}\ .
\end{equation}
We can find the electric charges via \eqref{eq:attractoreq},
\begin{align}\label{eq:em-t3}
\begin{split}
	Q_0 &= \frac1\Xi (\beta_0 +2 g^2\? (\beta_1)^2 (3 \beta_0 - \beta_1)) \ , \\ Q_1 =  Q_2 = Q_3 &=  \frac1\Xi (\beta_1 +2 g^2\? (\beta_1)^2 (\beta_0 + \beta_1))\ ,
\end{split}
\end{align}
with
\begin{equation}
	\Xi = 1 - 16 g^4\? \beta_0 (\beta_1)^3\ , \quad a = 2 \sqrt{2} g\? \sqrt{\beta_0 (\beta_1)^3}\ ,
\end{equation}
and the entropy via \eqref{eq:Bek-Haw-entropy},
\begin{align}\label{eq:S-t3}
\begin{split}
	S^{T^3} = & \frac{2 \pi}{\Xi} \Big( \beta_0 (\beta_1)^3 (1+16 g^4\? \beta_0 (\beta_1)^3) \\ &+ g^2\? (\beta_1)^4 (6 \beta_0 \beta_1 +3 (\beta_0)^2 - (\beta_1)^2) \Big)^{1/2}\ .
\end{split}
\end{align}
The entropy as a function of electric charges follows the expected behavior from the entropy function \eqref{eq:entropyfn}, where the angular momentum is also fixed to be
\begin{equation}\label{eq:J-stu}
\begin{split}
	{\cal J}^{T^3} =  \frac{2\? g\? (\beta_1)^2}{\Xi^2} & \Big( (3 \beta_0 + \beta_1) (1+16 g^4\? \beta_0 (\beta_1)^3) \\ &+ 8 g^2\? \beta_0 \beta_1 (\beta_0+3 \beta_1) \Big)\ .
\end{split}
\end{equation}
We have also checked explicitly the validity of \eqref{eq:entropyfn} for the full $STU$ model with four independent electric charges.

\subsection{The \texorpdfstring{$X^0 X^1$}{X0 X1} model with dyonic charges}\label{sec:x0x1sol}
Here we specialize to the so called $X^0 X^1$ model, for which the parameters in \eqref{eq:C-STU} are identified pairwise, $\beta_0 = \beta_2, \beta_1 = \beta_3$ such that
\begin{equation}\label{eq:C-X0X1}
	\cC^{\scriptscriptstyle X^0 X^1} = \{-\alpha, \alpha, -\alpha, \alpha; \beta_0, \beta_1, \beta_0, \beta_1 \}\ .
\end{equation}
This corresponds to having two independent electric charges and one free magnetic charge and thus generalizes the already existing purely electric solution of \cite{Cvetic:2005zi}. The set of conserved charges is given by
\begin{align}\label{eq:em-x0x1}
\begin{split}
	P \equiv P^0 = - P^1 =  -\frac{\alpha}{\Xi}\? (1+4 g^2 (\beta_0 \beta_1 - \alpha^2))\ ,\\ Q_{0, 1} =  \frac{\beta_{0,1}}{\Xi}\? (1+4 g^2 (\beta_0 \beta_1 - \alpha^2))\ ,
\end{split}
\end{align}
with 
\begin{equation}
\Xi = 1 - 16 g^4\? (\beta_0 \beta_1 - \alpha^2)^2\ , \quad a = 2 \sqrt{2} g\? (\beta_0 \beta_1 - \alpha^2)\ .
\end{equation}
The entropy and angular momentum can be computed via \eqref{eq:Bek-Haw-entropy} and \eqref{eq:J-gen}, assuming $Q_0\?Q_1 > P^2$,
\begin{align}\label{eq:S-x0x1}
S^{\scriptscriptstyle X^0 X^1} = & \frac{\pi}{4 g^2} \left(-1 + \sqrt{1 + 16 g^2\? (Q_0 Q_1 - P^2)}\ \right)\ , \nonumber \\              
% & = \frac{\pi\? a}{\sqrt{2} g\? (1 - \frac{a}{l}) }\ , \\
% {\cal J}^{\scriptscriptstyle X^0 X^1} = &\frac{a\? (Q_0 + Q_1)}{\sqrt{2} (1 - \frac{a}{l})} \ .
& = \frac{\pi\? {\cal J}^{\scriptscriptstyle X^0 X^1} }{2\? g\? (Q_0 + Q_1)} \ .
\end{align}
These quantities are again consistent with the entropy function \eqref{eq:entropyfn} and coincide with those of \cite{Cvetic:2005zi} in the limit of vanishing magnetic charge.

\section*{Acknowledgements}
We thank  B.\? Willett for useful correspondence and  S.M.\? Hosseini, A.\? Zaffaroni for discussions and for pointing out typos in previous versions of the paper.  KH is supported in part by the Bulgarian NSF grants DN08/3 and N28/5. SK is supported by the KU Leuven C1 grant ZKD1118 C16/16/005 and by the Belgian Federal Science Policy Office through the Inter-University Attraction Pole P7/37. CT is supported by the Agence Nationale de la Recherche (ANR) under the grant Black-dS-String (ANR-16-CE31-0004).

\bibliography{AdSrot}

\end{document}